# Theory of Slidetronics in Ferroelectric van der Waals Layers

Byeoksong Lee,[‡] Minki Lee,[‡] and Joongoo Kang[*]

Department of Physics and Chemistry, DGIST, Daegu 42988, Republic of Korea

Vertically stacked layers derived from non-ferroelectric monolayers offer a promising route to two-dimensional (2D) ferroelectrics, where polarization switching occurs via interlayer sliding at sub-unit cell scales. Here, we develop a theory of slidetronics based on the notion that sliding-induced switching $\boldsymbol{P} \longrightarrow \boldsymbol{P}'$ can also be achieved by applying an appropriate point-group operator $G$ to the entire system, such that $\boldsymbol{P}' = G\boldsymbol{P}$. Interlayer sliding and the transformation induced by the "generator" $G$ are thus equivalent in describing the relationship between the initial and final layer configurations. From this symmetry principle, we deduce that slidetronics can be classified by generators $G$; the generator $G$ must act as a symmetry operator for the constituent layers, while it is not a symmetry operator for the stacked layers as a whole; for a given 2D material, $G$ determines the interlayer sliding required for polarization switching; and sliding-induced complete polarization inversion is impossible in bilayers but can be realized in multilayers (e.g., $PdSe_2$ trilayers). These findings provide a framework for designing 2D ferroelectrics with targeted polarization-switching properties, as demonstrated through case studies of real materials.

**Introduction.** Interlayer sliding within two-dimensional (2D) van der Waals layers provides a unique symmetry-breaking mechanism for realizing 2D ferroelectricity (FE). Sliding-induced polarization switching [1–18] was first proposed by Li and Wu in 2017 [1] using bilayers of hexagonal boron nitride (h-BN) and transition metal dichalcogenides (TMDs) as model systems. The energy barriers for such lateral shifts are small, typically a few meV per unit cell [1,3], resulting in low critical electric fields for polarization switching. Recent experiments confirmed room-temperature FE in h-BN bilayers [8–10] and TMD bilayers and multilayers [11–18].

The crystal symmetry of 2D materials dictates which components of electric polarization are allowed and which are forbidden by symmetry constraints. In a recent pioneering work, Ji et al. [5] performed a comprehensive group theory analysis to identify all possible polar types for bilayer systems based on the 80 layer groups of constituent monolayers. They classified 2D systems into four polar types: in-plane polarization (IP), out-of-plane polarization (OP), combined polarization (CP) with both in-plane and out-of-plane components, and nonpolar (NP). For example, the group theory analysis showed that a bilayer composed of monolayers with layer group 55 cannot exhibit IP but can achieve other polar types (i.e., OP, CP, and NP) through specific stacking operations. This classification theory, while limited to bilayers, represents a milestone in 2D FE research.

However, when applied to the design of new 2D materials for *slidetronics*, this approach has limitations. While the previous study focuses on identifying possible polar types (IP, OP, CP, and NP) for a *given* bilayer system, slidetronics concerns the transformations between *pairs* of 2D layer configurations and the associated *changes* in polarization. Not all polar bilayer systems exhibit sliding FE, as this phenomenon relies on polarization induced by interlayer interactions. Furthermore, among the polarization components arising from interlayer interactions, only some are switchable through lateral shifts of the stacked layers. As a complementary theoretical framework, a theory of slidetronics should address the following critical questions:

(**1**) Which polarization components of a given layered system are switchable for sliding FE?

(**2**) How can these active components be switched through specific interlayer sliding motions?

(**3**) For a given structure, is there a definitive algorithm to identify all sliding-switchable phases with distinct electric polarizations?

(**4**) Is there a critical number of layers required for certain types of polarization switching?

(**5**) What are the design principles for 2D ferroelectrics with targeted polarization-switching properties?

Here, we address these fundamental questions (Q1–Q5) by developing a general theory of slidetronics, based on the notion that sliding-induced polarization switching can also occur *without* interlayer sliding. For any polarization switching in slidetronics, there always exists an appropriate point-group operator $G$ such that $\boldsymbol{P'} = G\boldsymbol{P}$. This implies that the sliding-induced structural transformation can be achieved by applying the operator $G$, referred to as a "generator", to the *entire* system. Conceptually, interlayer sliding and the transformation induced by $G$ are equivalent in describing the relationship between the initial and final phases of the transformation. From this symmetry principle, we deduce key properties of sliding FE: slidetronics can be classified by generators $G$; the generator $G$ must act as a symmetry operator for the constituent layers, while it is not a symmetry operator for the stacked layers as a whole; for a given 2D material, $G$ determines the interlayer sliding required for polarization transformation; and sliding-induced complete polarization inversion, corresponding to $G = I$ (inversion), is impossible in bilayer systems but can be realized in trilayer systems. These findings enable the design of materials with targeted polarization-switching properties. Key predictions include in-plane polarization switching in $As_2S_3/As_2Se_3$ heterobilayers, dipole locking in cellulose bilayers, and complete polarization inversion in $PdSe_2$ trilayers, all validated by density-functional theory (DFT) calculations.

**Symmetry principle.** For the sake of definiteness, we consider a bilayer $\mathbb{B}$ with an electric polarization $\boldsymbol{P} = \left(P_x, P_y, P_z\right)$ and want to transform it into $\boldsymbol{P'} = \left(P_x, -P_y, -P_z\right)$. A straightforward approach is to rotate the *entire* system $\mathbb{B}$ around the $x$-axis by 180° [Fig. 1(a)], such that $\mathbb{B} \longrightarrow C_{2x}\mathbb{B}$ and $\boldsymbol{P'} = C_{2x}\boldsymbol{P}$. At the core of sliding FE is the symmetry principle that if the bilayer $\mathbb{B}$ has the right crystal structure, the rotated bilayer $C_{2x}\mathbb{B}$ can be equivalently generated by laterally shifting one of the two parallel layers in $\mathbb{B}$ [Fig. 1(a)]. That is, $\mathbb{B} \longrightarrow T_{sld}\mathbb{B}$, where $T_{sld}$ is an operator that induces interlayer sliding at sub-unit cell scales. Since $C_{2x}\mathbb{B}$ and $T_{sld}\mathbb{B}$ represent two different perspectives of the same system, $T_{sld}\mathbb{B}$ must also exhibit the switched polarization $\boldsymbol{P}' = C_{2x}\boldsymbol{P}$. The bilayer systems $\mathbb{B}$ and $T_{sld}\mathbb{B}$ correspond to two stable phases on the free-energy surface in Fig. 1(a), exhibiting different polarizations, $\boldsymbol{P}$ and $C_{2x}\boldsymbol{P}$, respectively.

In the above example, the sliding-induced transformation, $\boldsymbol{P} \xrightarrow{T_{sld}} C_{2x}\boldsymbol{P}$, is "generated" by the rotation operator $C_{2x}$. More generally, other point-group operators can serve as a *generator $G$* of the polarization transformation $\boldsymbol{P} \xrightarrow{T_{sld}} G\boldsymbol{P}$. For a given $G$, the condition for sliding FE is given by

$$G\mathbb{B} = T(\boldsymbol{\tau})[T_{sld}\mathbb{B}], \qquad (1)$$

where $T(\boldsymbol{\tau})$ denotes a translation of the system $T_{sld}\mathbb{B}$ by a vector $\boldsymbol{\tau}$. To satisfy this condition, the "rotated" system $G\mathbb{B}$ must remain parallel to the original system $\mathbb{B}$. Consequently, for a vertically stacked system, the generator $G$ is restricted to the following set of point-group operators:

$$\{G\} = \{I, M_z, C_{nz}, S_{nz}, M_\alpha, C_{2\beta}\}, \qquad (2)$$

where $I$ represents inversion, and $M_z$ denotes a mirror reflection with respect to the plane at $z = 0$. The operator $C_{nz}$ ($S_{nz}$) represents an *n*-fold proper (improper) rotation around the z-axis. For a direction $\alpha$ (or $\beta$) lying in the *x-y* plane, $M_\alpha$ denotes a reflection across a plane normal to $\alpha$, while $C_{2\beta}$ denotes a two-fold rotation around an axis in the direction $\beta$.

Therefore, any polarization switching $\boldsymbol{P} \xrightarrow{T_{sld}} G\boldsymbol{P}$ in slidetronics can be labeled by a generator $G$ from the set in Eq. (2). In Fig. 1(b), we classify slidetronics into four types based on how sliding-induced changes in $\boldsymbol{P}$ occur. When $G = M_z$, only the out-of-plane polarization $P_z$ undergoes a sign change, while the in-plane polarization remains unaffected (type I). If $G \in \{M_\alpha, C_{nz}\}$, interlayer sliding affects only the in-plane polarization (type II). When $G \in \{C_{2\beta}, S_{nz}\}$, the out-of-plane and in-plane polarization components are interlocked (type III), allowing the in-plane polarization to be switched by an out-of-plane electric field and vice versa. Finally, if possible, choosing $G = I$ (inversion) would lead to a complete inversion of polarization induced by interlayer sliding (type IV). However, as we will show later, type-IV sliding FE is impossible in bilayer systems.

**Theory.** Building on the symmetry principle expressed in Eq. (1), we develop a general theoretical framework for slidetronics. To achieve this, we first define stacking operations for 2D layered systems. A bilayer system $\mathbb{B}$ is constructed by stacking a top monolayer $\mathbb{M}_t$ onto a bottom monolayer $\mathbb{M}_b$, with its position determined by an in-plane translation vector $\boldsymbol{\tau}_{xy}^t$ and an out-of-plane translation vector $\boldsymbol{\tau}_z^t$. The bilayer $\mathbb{B}$ can be represented as $\mathbb{B} = \mathbb{M}_b + T(\boldsymbol{\tau}_{xy}^t + \boldsymbol{\tau}_z^t)\mathbb{M}_t$. The translation operator $T$ satisfies $T(\boldsymbol{\tau}_1 + \boldsymbol{\tau}_2) = T(\boldsymbol{\tau}_1)T(\boldsymbol{\tau}_2)$ and thus commutes, i.e., $T(\boldsymbol{\tau}_1)T(\boldsymbol{\tau}_2) = T(\boldsymbol{\tau}_2)T(\boldsymbol{\tau}_1)$. In a homobilayer system, $\mathbb{M}_b$ and $\mathbb{M}_t$ can still differ. For example, given $\mathbb{M}_b = \mathbb{M}$, the top layer could be $\mathbb{M}_t = I\mathbb{M}$ or $\mathbb{M}_t = C_{4z}\mathbb{M}$. More generally, $\mathbb{M}_t = O_t\mathbb{M}$, where $O_t$ is a point-group operator that transforms $\mathbb{M}$ into $O_t\mathbb{M}$ while preserving its parallel orientation. If $O_t = E$ (the identity operator), then $\mathbb{M}_b = \mathbb{M}_t = \mathbb{M}$. For a bilayer $\mathbb{B}$, the sliding operator $T_{sld}$ in Eq. (1) is defined as $T_{sld}\mathbb{B} = \mathbb{M}_b + T(\boldsymbol{\tau}_{sld}^t)T(\boldsymbol{\tau}_{xy}^t + \boldsymbol{\tau}_z^t)\mathbb{M}_t$, where $T(\boldsymbol{\tau}_{sld}^t)$ acts solely on the top layer $\mathbb{M}_t$, with $\boldsymbol{\tau}_{sld}^t$ representing the interlayer sliding vector.

We present the symmetry conditions for the constituent monolayers $\mathbb{M}_b$ and $\mathbb{M}_t$ in a bilayer $\mathbb{B}$ and the equation for the interlayer sliding vector $\boldsymbol{\tau}_{sld}^t$ (see Methods for derivation). As illustrated in Fig. 1(c), the generator $G$ must act as a symmetry operator for the constituent layers, while it is not a symmetry operator for the stacked layers as a whole. Equation (1) has two possible solutions, depending on whether $G$ flips the sign of the $z$-component $P_z$ in the transformation $\boldsymbol{P} \xrightarrow{T_{sld}} G\boldsymbol{P}$. Thus, we divide the set $\{G\}$ in Eq. (2) into two categories, $G^+ \in \{M_\alpha, C_{nz}\}$ and $G^- \in \{I, C_{2\beta}, M_z, S_{nz}\}$, according to their transformation properties: $G^+\hat{\boldsymbol{z}} = \hat{\boldsymbol{z}}$ and $G^-\hat{\boldsymbol{z}} = -\hat{\boldsymbol{z}}$ for the unit vector $\hat{\boldsymbol{z}}$.

The necessary conditions for sliding FE in a bilayer $\mathbb{B}$ are summarized as follows:

**(i) Case of $\boldsymbol{G^+} \in \{\boldsymbol{M_\alpha}, \boldsymbol{C_{nz}}\}$:** The generator $G^+$ of polarization transformation must also act as a symmetry operator of the constituent monolayers in the bilayer $\mathbb{B}$:

$$G^+\mathbb{M}_b = T(\boldsymbol{\tau}^{bb})\mathbb{M}_b \text{ and } G^+\mathbb{M}_t = T(\boldsymbol{\tau}^{tt})\mathbb{M}_t \qquad (3a)$$

for translation vectors $\boldsymbol{\tau}^{bb}$ and $\boldsymbol{\tau}^{tt}$, respectively. The sliding vector $\boldsymbol{\tau}^t_{sld}$ in $\mathbb{B} \longrightarrow T_{sld}\mathbb{B}$ is given as

$$\boldsymbol{\tau}^t_{sld} = G^+\boldsymbol{\tau}^t_{xy} - \boldsymbol{\tau}^t_{xy} + \boldsymbol{\tau}^{tt} - \boldsymbol{\tau}^{bb} \bmod (\boldsymbol{a}_1, \boldsymbol{a}_2), \qquad (3b)$$

which is determined only up to integer multiplies of the lattice vectors $\boldsymbol{a}_1$ and $\boldsymbol{a}_2$ of the 2D system.

**(ii) Case of $\boldsymbol{G^-} \in \{\boldsymbol{I}, \boldsymbol{C_{2\beta}}, \boldsymbol{M_z}, \boldsymbol{S_{nz}}\}$:** For a generator $G^-$, equation (1) implies that $G^-$ converts the bottom layer $\mathbb{M}_b$ into the top layer $\mathbb{M}_t$, and vice versa,

$$G^-\mathbb{M}_b = T(\boldsymbol{\tau}^{bt})\mathbb{M}_t \text{ and } G^-\mathbb{M}_t = T(\boldsymbol{\tau}^{tb})\mathbb{M}_b \qquad (4a)$$

for translation vectors $\boldsymbol{\tau}^{bt}$ and $\boldsymbol{\tau}^{tb}$, respectively. The interlayer sliding vector $\boldsymbol{\tau}^t_{sld}$ is then given by

$$\boldsymbol{\tau}^t_{sld} = -G^-\boldsymbol{\tau}^t_{xy} - \boldsymbol{\tau}^t_{xy} + \boldsymbol{\tau}^{bt} - \boldsymbol{\tau}^{tb} \bmod (\boldsymbol{a}_1, \boldsymbol{a}_2). \qquad (4b)$$

There are two things to note on $\boldsymbol{\tau}^t_{sld}$ in Eqs. (3b) and (4b). If equation (1) holds with $\boldsymbol{\tau}^t_{sld} = \boldsymbol{0}$ mod $(\boldsymbol{a}_1, \boldsymbol{a}_2)$, no interlayer sliding occurs, meaning that the transformation no longer describes sliding FE. Instead, the operator $G$ serves as a symmetry operator for the bilayer $\mathbb{B}$ as a whole, restricting the polarization $\boldsymbol{P}$ through the relation $G\boldsymbol{P} = \boldsymbol{P}$. Additionally, the sliding vector $\boldsymbol{\tau}^t_{sld}$ remains invariant under changes in the origin of point-group operators (such as $G$ and the $O_t$ of the stacking operation), while all other translation vectors depend on the choice of the origin.

As a simple example, we consider a bilayer of h-BN. For h-BN monolayers $\mathbb{M}$, the in-plane translation vector in the bilayer $\mathbb{B} = \mathbb{M} + T(\boldsymbol{\tau}^t_{xy} + \boldsymbol{\tau}^t_z)\mathbb{M}$ is given by $\boldsymbol{\tau}^t_{xy} = -\boldsymbol{\tau}_{BN}$, where $\boldsymbol{\tau}_{BN}$ is a vector that points along a B–N bond in $\mathbb{M}$. For $G^- = M_z$, equation (4a) is satisfied since $M_z$ is a symmetry operator for the monolayer $\mathbb{M}$. According to Eq. (4b), we obtain $\boldsymbol{\tau}^t_{sld} = M_z\boldsymbol{\tau}_{BN} + \boldsymbol{\tau}_{BN} = 2\boldsymbol{\tau}_{BN}$. Since $\boldsymbol{\tau}^t_{sld} \neq \boldsymbol{0}$ mod $(\boldsymbol{a}_1, \boldsymbol{a}_2)$, type-I sliding FE is realized with $G^- = M_z$, resulting in $P_z \xrightarrow{M_z} -P_z$. However, the h-BN bilayer cannot support other types of slidetronics. For instance, to achieve type-II sliding FE, we could choose $G^+ = C_{3v}$, which satisfies Eq. (3a). From Eq. (3b), the corresponding sliding vector would be $\boldsymbol{\tau}^t_{sld} = -C_{3v}\boldsymbol{\tau}_{BN} + \boldsymbol{\tau}_{BN}$, which is equal to a lattice vector of $\mathbb{M}$, violating the necessary condition $\boldsymbol{\tau}^t_{sld} \neq \boldsymbol{0}$ mod $(\boldsymbol{a}_1, \boldsymbol{a}_2)$ for sliding FE.

The symmetry principle developed for bilayer systems $\mathbb{B}$ can be extended to trilayer systems $\mathbb{T}$. The conditions for sliding FE in trilayers, derived from the symmetry relation $G\mathbb{T} = T(\boldsymbol{\tau})[T_{sld}\mathbb{T}]$, are discussed in Methods (see 'Theory for trilayer systems').

**Applications.** The results presented in Eqs. (1)–(4) provide a theoretical framework for understanding sliding FE in 2D layered materials. The applications of our theory are two-fold.

First, for a given layered system, it enables the automatic identification of all sliding-switchable phases with distinct polarizations, which is particularly useful for high-throughput searches of sliding ferroelectrics. A definitive algorithm, described below, reveals the structure–property relationships and addresses questions Q1–Q3 posed in the Introduction.

Second, our theory offers practical guidance for designing new sliding ferroelectrics with targeted polarization-switching properties. For instance, we mathematically prove that complete polarization inversion (type IV) is impossible in bilayer systems but can occur in trilayer systems (Q4). Finally, we present three case studies demonstrating how our findings facilitate the design of novel sliding ferroelectrics with desired switching properties (Q5).

Starting with a stable bilayer phase, we identify all other equally stable, sliding-switchable phases with distinct electric polarizations through the following steps (see Supplemental Section 1 [19] for the discussion on trilayer systems):

**1. Generator selection:** Consider a homobilayer phase $\mathbb{B} = \mathbb{M} + T(\boldsymbol{\tau}_{xy}^t + \boldsymbol{\tau}_z^t)[O_t\mathbb{M}]$, where $\mathbb{M}_b = \mathbb{M}$ (bottom layer) and $\mathbb{M}_t = O_t\mathbb{M}$ (top layer). From the set $\{G\}$ in Eq. (2), we select all possible generators $G$ that satisfy the conditions in Eqs. (3a) and (4a). Specifically, if $G = G^+$, both $G^+$ and $O_t^{-1}G^+O_t$ must be symmetry operators of the monolayer $\mathbb{M}$. If $G = G^-$, both $G^-O_t$ and $O_t^{-1}G^-$ must act as symmetry operators of $\mathbb{M}$.

**2. Sliding vector calculation:** For each selected generator $G$, we calculate the corresponding interlayer sliding vector $\boldsymbol{\tau}_{sld}^t(G)$ using Eqs. (3b) and (4b).

**3. Final generator selection:** We retain only those generators for which $\boldsymbol{\tau}_{sld}^t(G) \neq \mathbf{0} \bmod (\boldsymbol{a}_1, \boldsymbol{a}_2)$. If multiple generators have the same interlayer sliding vector, i.e., $\boldsymbol{\tau}_{sld}^t(G) = \boldsymbol{\tau}_{sld}^t(G') \bmod (\boldsymbol{a}_1, \boldsymbol{a}_2)$, we keep only one of them.

**4. Polarization transformation:** Each remaining generator $G$ corresponds to a bilayer phase $T_{sld}(G)\mathbb{B}$ with a transformed polarization $\boldsymbol{P}' = G\boldsymbol{P}$.

**Proof of the impossibility of type-IV slidetronics in bilayer systems.** While bilayer systems can exhibit types I, II, and III sliding FE through appropriate selection of constituent monolayers and stacking operations, type-IV (dipole inversion) is fundamentally impossible. To prove this, we apply the dipole-inversion generator $G^- = I$ to both sides of the second relation in Eq. (4a), yielding $G^-G^-\mathbb{M}_t = G^-[T(\boldsymbol{\tau}^{tb})\mathbb{M}_b] = T(G^-\boldsymbol{\tau}^{tb})[G^-\mathbb{M}_b]$, where the last step follows from the general rule $G[T(\boldsymbol{\tau})\mathbb{L}] = T(G\boldsymbol{\tau})[G\mathbb{L}]$ for a layered system $\mathbb{L}$. Given that $G^-\mathbb{M}_b = T(\boldsymbol{\tau}^{bt})\mathbb{M}_t$ in Eq. (4a), we obtain $G^-G^-\mathbb{M}_t = T(G^-\boldsymbol{\tau}^{tb} + \boldsymbol{\tau}^{bt})\mathbb{M}_t$. Since $G^-G^- = E$ and $G^-\boldsymbol{\tau}^{tb} = -\boldsymbol{\tau}^{tb}$ for $G^- = I$, it follows that $\mathbb{M}_t = T(-\boldsymbol{\tau}^{tb} + \boldsymbol{\tau}^{bt})\mathbb{M}_t$, meaning that $\boldsymbol{\tau}^{tb} = \boldsymbol{\tau}^{bt} \bmod (\boldsymbol{a}_1, \boldsymbol{a}_2)$. Consequently, with $G^- = I$, the interlayer sliding vector $\boldsymbol{\tau}_{sld}^t$ in Eq. (4b) must always be zero. In other words, with nonzero $\boldsymbol{\tau}_{sld}^t$, equation (1) has no solution for bilayer systems $\mathbb{B}$. In contrast, type-IV sliding FE is achievable in trilayer systems (see Fig. 4(a) and 'Case study 3' below).

In the following, we present case studies demonstrating how different types of slidetronics can be realized by designing specific 2D stacked bilayers or trilayers.

**Case study 1: Type-III slidetronics in a cellulose bilayer.** As summarized in Fig. 1(b), type-III sliding FE is achievable with a generator $G^- \in \{C_{2\beta}, S_{nz}\}$. Here we choose $G^- = C_{2x}$. Equation (4a) implies that the targeted switching property (dipole locking between $P_y$ and $P_z$) can be realized by vertically stacking two monolayers $\mathbb{M}$ with $C_{2x}$ symmetry (i.e., $\mathbb{M}_t = \mathbb{M}_b = \mathbb{M}$). As a candidate for $\mathbb{M}$, we consider a cellulose sheet [20,21], whose atomic structure remains unchanged when first rotated 180° around the $x$-axis and then translated by a fraction of the lattice period $\boldsymbol{\tau} = \frac{\boldsymbol{a}_1}{2}$ [Fig. 2(a)]; that is, $\mathbb{M} = T\left(\frac{\boldsymbol{a}_1}{2}\right)[C_{2x}\mathbb{M}]$. This $C_{2x}$ symmetry restricts the monolayer polarization $\boldsymbol{P}$ to a nonzero $P_x$ component ($P_y = P_z = 0$). However, in a cellulose bilayer [Fig. 2(b)], the $y$ and $z$ components of the bilayer polarization can emerge due to interlayer interactions.

Figure 2(c) shows the DFT total energy of the bilayer $\mathbb{B} = \mathbb{M} + T(\boldsymbol{\tau}_{xy}^t + \boldsymbol{\tau}_z^t)\mathbb{M}$ as a function of the translation vector $\boldsymbol{\tau}_{xy}^t$ (Methods and [22–28]). Two low-energy phases, marked by open circles, were identified. We start with the phase in the right half-plane at $\boldsymbol{\tau}_{xy}^t = \frac{1+\varepsilon}{4}\boldsymbol{a}_1 + 0.57\boldsymbol{a}_2$, where $\varepsilon = 0.05$ denotes a small deviation from $\frac{1}{4}$ in the coefficient of $\boldsymbol{a}_1$. Since $G^- = C_{2x}$ and $\boldsymbol{\tau}^{bt} = \boldsymbol{\tau}^{tb} = \frac{\boldsymbol{a}_1}{2}$, it follows from Eq. (4b) that $\boldsymbol{\tau}_{sld}^t = -C_{2x}\boldsymbol{\tau}_{xy}^t - \boldsymbol{\tau}_{xy}^t = -\frac{1+\varepsilon}{2}\boldsymbol{a}_1$ mod $(\boldsymbol{a}_1, \boldsymbol{a}_2)$. Hence, type-III sliding FE can be achieved with $\boldsymbol{\tau}_{sld}^t = -\frac{1+\varepsilon}{2}\boldsymbol{a}_1$ (path 1) or $\boldsymbol{\tau}_{sld}^t = \frac{1-\varepsilon}{2}\boldsymbol{a}_1$ (path 2). The energy barriers are $E_b = 3.2$ meV/Å$^2$ along path 1 and 4.1 meV/Å$^2$ along path 2.

Using the Berry-phase theory of polarization [29–32], we calculate the bilayer polarization $\boldsymbol{P}$ as a function of the translation vector $\boldsymbol{\tau}_{xy}^t$ [Fig. 2(c)]. Since $P_x$ is not switchable (and thus cannot be determined), we instead plot the change in $P_x$ induced by interlayer interactions. The in-plane polarizations are represented by arrows, while the out-of-plane component $P_z$ is indicated by the color of the arrows. As the stacking order shifts from $\boldsymbol{\tau}_{xy}^t = \frac{1+\varepsilon}{4}\boldsymbol{a}_1 + 0.57\boldsymbol{a}_2$ to $\boldsymbol{\tau}_{xy}^t = -\frac{1+\varepsilon}{4}\boldsymbol{a}_1 + 0.57\boldsymbol{a}_2$, both $P_y$ and $P_z$ reverse their signs, while $P_x$ remains unchanged. Since $P_y$ is interlocked with $P_z$, the in-plane polarization $P_y$ can be switched by applying an out-of-plane electric field, and vice versa, enabling applications such as switchable ferroelectric diodes [33].

**Case study 2: Type-II slidetronics in an As₂S₃ bilayer.** The next target property is type-II sliding FE, which is achievable with a generator $G^+ \in \{M_\alpha, C_{nz}\}$ [Fig. 1(b)]. Here, we choose $G^+ = M_y$ and consider an As₂S₃ bilayer [34,35] because, as required by Eq. (3a), its constituent monolayer $\mathbb{M}$ has $M_y$ as one of its symmetry operators $S \in \{M_y, M_z, C_{2x}\}$. When the origin is chosen as shown in Fig. 3(a), the translation vector $\boldsymbol{\tau}$ in $S\mathbb{M} = T(\boldsymbol{\tau})\mathbb{M}$ is $\boldsymbol{\tau} = \boldsymbol{0}$ for $S = M_y$, and $\boldsymbol{\tau} = \frac{\boldsymbol{a}_2}{2}$ for $S = M_z$ and $C_{2x}$.

We consider a bilayer configuration $\mathbb{B} = \mathbb{M} + T(\boldsymbol{\tau}_{xy}^t + \boldsymbol{\tau}_z^t)[I\mathbb{M}]$, where $\mathbb{M}_b = \mathbb{M}$ and $\mathbb{M}_t = I\mathbb{M}$, obtained by exfoliating bulk As₂S₃ [34–36]. (See Supplemental Sections 2 and 3 for results on a different configuration where $\mathbb{M}_t = \mathbb{M}$.) However, this bilayer $\mathbb{B}$ possesses inversion symmetry regardless of the choice of $\boldsymbol{\tau}_{xy}^t$. (Proof: For $G^- = I$ (inversion), equation (4a) is satisfied with $\boldsymbol{\tau}^{bt} = \boldsymbol{\tau}^{tb}$, and the sliding vector $\boldsymbol{\tau}_{sld}^t$ in Eq. (4b) becomes $\boldsymbol{\tau}_{sld}^t = -I\boldsymbol{\tau}_{xy}^t - \boldsymbol{\tau}_{xy}^t = \boldsymbol{0}$, indicating that $I\boldsymbol{P} = \boldsymbol{P}$ and thus $\boldsymbol{P} = \boldsymbol{0}$.) Hence, the As$_2$S$_3$ bilayer $\mathbb{B}$ is not a sliding FE material.

One way to overcome this limitation is to form a heterobilayer composed of As$_2$S$_3$ ($\mathbb{M}_{\text{S}}$) and As$_2$Se$_3$ ($\mathbb{M}_{\text{Se}}$). In the heterobilayer $\mathbb{B}_{\text{S/Se}} = \mathbb{M}_{\text{S}} + T(\boldsymbol{\tau}_{xy}^t + \boldsymbol{\tau}_z^t)[I\mathbb{M}_{\text{Se}}]$, the inversion symmetry of the homobilayer $\mathbb{B}$ is broken, leading to $\boldsymbol{P} = (-0.22, 0.27, 0.85)\ pC/m$ in DFT calculations. We assume a lattice-matched As$_2$S$_3$/As$_2$Se$_3$ bilayer with the lowest-energy stacking order shown in the left panel of Fig. 3(c). This heterobilayer $\mathbb{B}_{\text{S/Se}}$ now supports type-II sliding FE, generated by $G^+ = M_y$ through top-layer sliding with $\boldsymbol{\tau}_{sld}^t = -2\boldsymbol{\tau}_y^t$ along the $y$-axis [Fig. 3(b)]; i.e., $(P_x, P_y, P_z) \xrightarrow{M_y} (P_x, -P_y, P_z)$. The energy barrier for this transformation is $E_b = 5.6\ \text{meV/Å}^2$.

Another way to break the inversion symmetry of the homobilayer $\mathbb{B}$ is to apply an external electric field $\mathcal{E}_z$ along the vertical direction. As shown in the right panel of Fig. 3(c), in-plane polarization is induced by the out-of-plane field $\mathcal{E}_z$. The components $P_x(\mathcal{E}_z)$ and $P_y(\mathcal{E}_z)$ are odd functions of $\mathcal{E}_z$. When top-layer sliding occurs with $\boldsymbol{\tau}_{sld}^t = -2\boldsymbol{\tau}_y^t$ (corresponding to $G^+ = M_y$) at a constant $\mathcal{E}_z$, the sign of $P_y$ flips, while $P_x$ remains unchanged. The energy barrier for the interlayer sliding is $E_b = 5.0\ \text{meV/Å}^2$ at $\mathcal{E}_z = 0.25\ \text{V/Å}$.

**Case study 3: Type-IV slidetronics in a PdSe$_2$ trilayer.** To achieve type-IV sliding FE in a homo-trilayer $\mathbb{T}$ with the generator $G^- = I$, the middle layer $\mathbb{M}_m$ in $\mathbb{T}$ should possess inversion symmetry, as required by Eq. (8a) in Methods [Fig. 1(c)]. Here, we consider a PdSe$_2$ monolayer $\mathbb{M}$ [37,38], which is a pentagonal phase having identity $E$ and inversion $I$ as symmetry operators [Fig. 4(b)]. The trilayer configuration is designed as $\mathbb{T} = \mathbb{M} + T(\boldsymbol{\tau}_{xy}^m + \boldsymbol{\tau}_z^m)[C_{4z}\mathbb{M}] + T(\boldsymbol{\tau}_{xy}^t + \boldsymbol{\tau}_z^t)\mathbb{M}$, where $\mathbb{M}_b = \mathbb{M}_t = \mathbb{M}$ and the middle layer $\mathbb{M}_m = C_{4z}\mathbb{M}$. Although the aspect ratio of the PdSe$_2$ unit cell slightly deviates from one ($a_y/a_x = 1.03$), we assume a square lattice for $\mathbb{M}$ in our DFT calculations to maintain lattice matching between the layers. With a Pd atom in $\mathbb{M}$ as the inversion center, we have $I\mathbb{M} = \mathbb{M}$ and $I(C_{4z}\mathbb{M}) = C_{4z}\mathbb{M}$, leading to $\boldsymbol{\tau}^{bt} = \boldsymbol{\tau}^{tb} = \boldsymbol{\tau}^{mm} = \boldsymbol{0}$ in Eq. (8a). From Eqs. (8b) and (8c), the sliding vectors are given by $\boldsymbol{\tau}_{sld}^t = -I\boldsymbol{\tau}_{xy}^t - \boldsymbol{\tau}_{xy}^t = \boldsymbol{0}$ for the top layer and $\boldsymbol{\tau}_{sld}^m = I\boldsymbol{\tau}_{xy}^m - \boldsymbol{\tau}_{xy}^m - I\boldsymbol{\tau}_{xy}^t = -2\boldsymbol{\tau}_{xy}^m + \boldsymbol{\tau}_{xy}^t$ for the middle layer. The trilayer $\mathbb{T}$, shown in the left panel of Fig. 4(c), has a polarization $\boldsymbol{P} = (-3.91, 0.42, 1.26)\ pC/m$, while the laterally shifted trilayer $T_{sld}\mathbb{T}$, shown in the right panel, exhibits an inverted polarization $-\boldsymbol{P}$.

In summary, starting from the symmetry relation $G\mathbb{L} = T(\boldsymbol{\tau})[T_{sld}\mathbb{L}]$ for a layered system $\mathbb{L}$ (e.g., bilayer $\mathbb{B}$ and trilayer $\mathbb{T}$), we deduce key properties of sliding FE: (1) slidetronics can be classified by generators $G$; (2) the generator $G$ must act as a symmetry operator for the constituent layers, while it is not a symmetry operator for the stacked layers as a whole; (3) for a given 2D material, $G$ determines the interlayer sliding required for polarization transformation; (4) and a minimum of three layers is required for sliding-induced dipole inversion. Given a 2D layered system, a definitive algorithm is devised for the automatic identification of all sliding-switchable phases with distinct polarizations, which will be particularly useful for high-throughput searches of novel sliding ferroelectrics. Finally, we present three case studies—in-plane polarization switching in $As_2S_3/As_2Se_3$ heterobilayers, dipole locking in cellulose bilayers, and complete polarization inversion in $PdSe_2$ trilayers—that illustrate how targeted polarization switching can be achieved through the theory-guided design of 2D layered materials.

## Methods

**Theory for bilayer systems.** In the symmetry principle expressed in Eq. (1), the left-hand side, $G\mathbb{B}$, represents the bilayer $\mathbb{B} = \mathbb{M}_b + T(\boldsymbol{\tau}_{xy}^t + \boldsymbol{\tau}_z^t)\mathbb{M}_t$ transformed by the generator $G$, i.e., $G\mathbb{B} = G\mathbb{M}_b + G[T(\boldsymbol{\tau}_{xy}^t + \boldsymbol{\tau}_z^t)\mathbb{M}_t]$. Using the general rule $G[T(\boldsymbol{\tau})\mathbb{L}] = T(G\boldsymbol{\tau})[G\mathbb{L}]$, which holds for any layered system $\mathbb{L}$, we can rewrite this as $G\mathbb{B} = G\mathbb{M}_b + T(G\boldsymbol{\tau}_{xy}^t + G\boldsymbol{\tau}_z^t)[G\mathbb{M}_t]$. As discussed in the main text, we classify the generator $G$ into two categories, $G^+ \in \{M_\alpha, C_{nz}\}$ and $G^- \in \{I, C_{2\beta}, M_z, S_{nz}\}$, depending on whether the operator $G$ alters the sign of $\boldsymbol{\tau}_z$, i.e., $G^\pm\boldsymbol{\tau}_z = \pm\boldsymbol{\tau}_z$. With this classification, $G^\pm\mathbb{B}$ can be expressed as

$$G^\pm\mathbb{B} = G^\pm\mathbb{M}_b + T(G^\pm\boldsymbol{\tau}_{xy}^t \pm \boldsymbol{\tau}_z^t)[G^\pm\mathbb{M}_t]. \qquad (5)$$

Given that $T_{sld}\mathbb{B} = \mathbb{M}_b + T(\boldsymbol{\tau}_{sld}^t)T(\boldsymbol{\tau}_{xy}^t + \boldsymbol{\tau}_z^t)\mathbb{M}_t$ and using the property $T(\boldsymbol{\tau}_1)T(\boldsymbol{\tau}_2) = T(\boldsymbol{\tau}_1 + \boldsymbol{\tau}_2)$, the right-hand side of Eq. (1) can be expressed as

$$T(\boldsymbol{\tau})[T_{sld}\mathbb{B}] = T(\boldsymbol{\tau})\mathbb{M}_b + T(\boldsymbol{\tau} + \boldsymbol{\tau}_{sld}^t + \boldsymbol{\tau}_{xy}^t + \boldsymbol{\tau}_z^t)\mathbb{M}_t. \qquad (6)$$

The symmetry principle, $G^{\pm}\mathbb{B} = T(\boldsymbol{\tau})[T_{sld}\mathbb{B}]$, in Eq. (1) must hold for any vertical translation vector $\boldsymbol{\tau}_z^t$. To satisfy this condition, the generator $G$ must act as a symmetry operator for the constituent layers, while it does not serve as a symmetry operator for the stacked layers as a whole [Fig. 1(c)]. Equation (1) has two possible solutions:

**(i) Case of $\boldsymbol{G^+} \in \{\boldsymbol{M_\alpha, C_{nz}}\}$:** By equating the right-hand sides of Eqs. (5) and (6) for a generator $G^+$, we obtain $G^+\mathbb{M}_b = T(\boldsymbol{\tau})\mathbb{M}_b$ and $G^+\mathbb{M}_t = T(\boldsymbol{\tau} + \boldsymbol{\tau}_{sld}^t + \boldsymbol{\tau}_{xy}^t - G^+\boldsymbol{\tau}_{xy}^t)\mathbb{M}_t$. This indicates that when applied to individual monolayers, $G^+$ must act as a symmetry operator. Therefore, the symmetry condition for the constituent monolayers of $\mathbb{B}$ is given by Eq. (3a), rewritten as $G^+\mathbb{M}_b = T(\boldsymbol{\tau}^{bb})\mathbb{M}_b$ and $G^+\mathbb{M}_t = T(\boldsymbol{\tau}^{tt})\mathbb{M}_t$. Combining these equations, we obtain the equations for the translation vectors: $\boldsymbol{\tau} = \boldsymbol{\tau}^{bb}$ and $\boldsymbol{\tau}^{tt} = \boldsymbol{\tau} + \boldsymbol{\tau}_{sld}^t + \boldsymbol{\tau}_{xy}^t - G^+\boldsymbol{\tau}_{xy}^t$. Hence, the interlayer sliding vector is given by $\boldsymbol{\tau}_{sld}^t = G^+\boldsymbol{\tau}_{xy}^t - \boldsymbol{\tau}_{xy}^t + \boldsymbol{\tau}^{tt} - \boldsymbol{\tau}^{bb}$ mod $(\boldsymbol{a_1, a_2})$, as given in Eq. (3b).

**(ii) Case of $\boldsymbol{G^-} \in \{\boldsymbol{I, C_{2\beta}, M_z, S_{nz}}\}$:** By equating the right-hand sides of Eqs. (5) and (6) for a generator $G^-$, we obtain the relations $G^-\mathbb{M}_b = T(\boldsymbol{\tau} + \boldsymbol{\tau}_{sld}^t + \boldsymbol{\tau}_{xy}^t + \boldsymbol{\tau}_z^t)\mathbb{M}_t$ and $G^-\mathbb{M}_t = T(\boldsymbol{\tau} - G^-\boldsymbol{\tau}_{xy}^t + \boldsymbol{\tau}_z^t)\mathbb{M}_b$. This implies that $G^-$ interchanges the bottom layer $\mathbb{M}_b$ and the top layer $\mathbb{M}_t$. The symmetry condition for the constituent monolayers is thus given by Eq. (4a), rewritten as $G^-\mathbb{M}_b = T(\boldsymbol{\tau}^{bt})\mathbb{M}_t$ and $G^-\mathbb{M}_t = T(\boldsymbol{\tau}^{tb})\mathbb{M}_b$. Combining these equations, we derive the equations for the translation vectors: $\boldsymbol{\tau}^{bt} = \boldsymbol{\tau} + \boldsymbol{\tau}_{sld}^t + \boldsymbol{\tau}_{xy}^t + \boldsymbol{\tau}_z^t$ and $\boldsymbol{\tau}^{tb} = \boldsymbol{\tau} - G^-\boldsymbol{\tau}_{xy}^t + \boldsymbol{\tau}_z^t$. By solving these equations for the unknown $\boldsymbol{\tau}$ and $\boldsymbol{\tau}_{sld}^t$, we obtain the sliding vector $\boldsymbol{\tau}_{sld}^t = -G^-\boldsymbol{\tau}_{xy}^t - \boldsymbol{\tau}_{xy}^t + \boldsymbol{\tau}^{bt} - \boldsymbol{\tau}^{tb}$ mod $(\boldsymbol{a_1, a_2})$, as given in Eq. (4b).

**Theory for trilayer systems.** We consider a trilayer system, $\mathbb{T} = \mathbb{M}_b + T(\boldsymbol{\tau}_{xy}^m + \boldsymbol{\tau}_z^m)\mathbb{M}_m + T(\boldsymbol{\tau}_{xy}^t + \boldsymbol{\tau}_z^t)\mathbb{M}_t$, which consists of the bottom (b), middle (m), and top (t) monolayers. The laterally shifted trilayer is defined as $T_{sld}\mathbb{T} = \mathbb{M}_b + T(\boldsymbol{\tau}_{sld}^m)T(\boldsymbol{\tau}_{xy}^m + \boldsymbol{\tau}_z^m)\mathbb{M}_m + T(\boldsymbol{\tau}_{sld}^t)T(\boldsymbol{\tau}_{xy}^t + \boldsymbol{\tau}_z^t)\mathbb{M}_t$.

The general principle developed for bilayer systems can be extended to trilayer systems to derive the conditions for sliding FE in trilayers, as summarized below:

**(i) Case of $G^+ \in \{M_\alpha, C_{nz}\}$:** For $G = G^+$, the condition for sliding FE, $G^+\mathbb{T} = T(\tau)[T_{sld}\mathbb{T}]$, implies that

$$G^+\mathbb{M}_b = T(\tau^{bb})\mathbb{M}_b, \; G^+\mathbb{M}_m = T(\tau^{mm})\mathbb{M}_m, \; G^+\mathbb{M}_t = T(\tau^{tt})\mathbb{M}_t, \quad (7a)$$

which indicates that $G^+$ should act as a symmetry operator for individual monolayers. The interlayer sliding vectors are then derived as

$$\tau^m_{sld} = G^+\tau^m_{xy} - \tau^m_{xy} + \tau^{mm} - \tau^{bb}, \quad\quad (7b)$$

$$\tau^t_{sld} = G^+\tau^t_{xy} - \tau^t_{xy} + \tau^{tt} - \tau^{bb}. \quad\quad (7c)$$

**(ii) Case of $G^- \in \{I, C_{2\beta}, M_z, S_{nz}\}$:** For $G = G^-$, the symmetry relation $G^-\mathbb{T} = T(\tau)[T_{sld}\mathbb{T}]$ is satisfied only when

$$G^-\mathbb{M}_b = T(\tau^{bt})\mathbb{M}_t, \; G^-\mathbb{M}_m = T(\tau^{mm})\mathbb{M}_m, \; G^-\mathbb{M}_t = T(\tau^{tb})\mathbb{M}_b. \quad (8a)$$

Here, the generator $G^-$ acts as a symmetry operator for the middle layer $\mathbb{M}_m$, while it converts the bottom layer $\mathbb{M}_b$ into the top layer $\mathbb{M}_t$, and vice versa. The sliding vectors in this case are

$$\tau^m_{sld} = G^-\tau^m_{xy} - \tau^m_{xy} - G^-\tau^t_{xy} + \tau^{mm} - \tau^{tb}, \quad (8b)$$

$$\tau^t_{sld} = -G^-\tau^t_{xy} - \tau^t_{xy} + \tau^{bt} - \tau^{tb}. \quad\quad (8c)$$

**DFT calculations.** The total energies and electronic structures of 2D layered materials were calculated using DFT within the generalized gradient approximation (GGA-PBE [22]), as implemented in the Vienna ab-initio Simulation Package (VASP [39,40]). To account for van der Waals interactions, we employed Grimme's DFT-D3 method [41] for 2D systems derived from $As_2S_3$, $As_2Se_3$, and $PdSe_2$ layers. It was shown that the PBE exchange-correlation potential, combined with Grimme's DFT-D2 method [42], accurately predicts the cell parameters of bulk cellulose phases and the stability of cellulose polymorphs [43]. As in our previous work [21] on cellulose nanocrystals, we thus used the PBE-D2 method for cellulose bilayers. Our DFT calculations utilized the projector augmented wave (PAW) method [27] with an energy cutoff of 400 eV for the plane-wave part of the wave function. All atomic structures were fully relaxed until residual forces were reduced below 0.05 eV/Å. The nudged elastic band (NEB) method [28] was employed to determine the energy barriers associated with interlayer sliding motions. Electric polarization was calculated using the Berry-phase theory of polarization [29–32]. A vacuum size of 100 Å was applied to ensure accurate calculations of the $z$-component polarization values in the supercell simulations of 2D materials.

## Acknowledgements

This work was supported by the National Research Foundation grant (No. NRF-2022M3D1A1026816) funded by the Ministry of Science and ICT of Korea.

‡These authors contributed equally.

* joongoo.kang@dgist.ac.kr

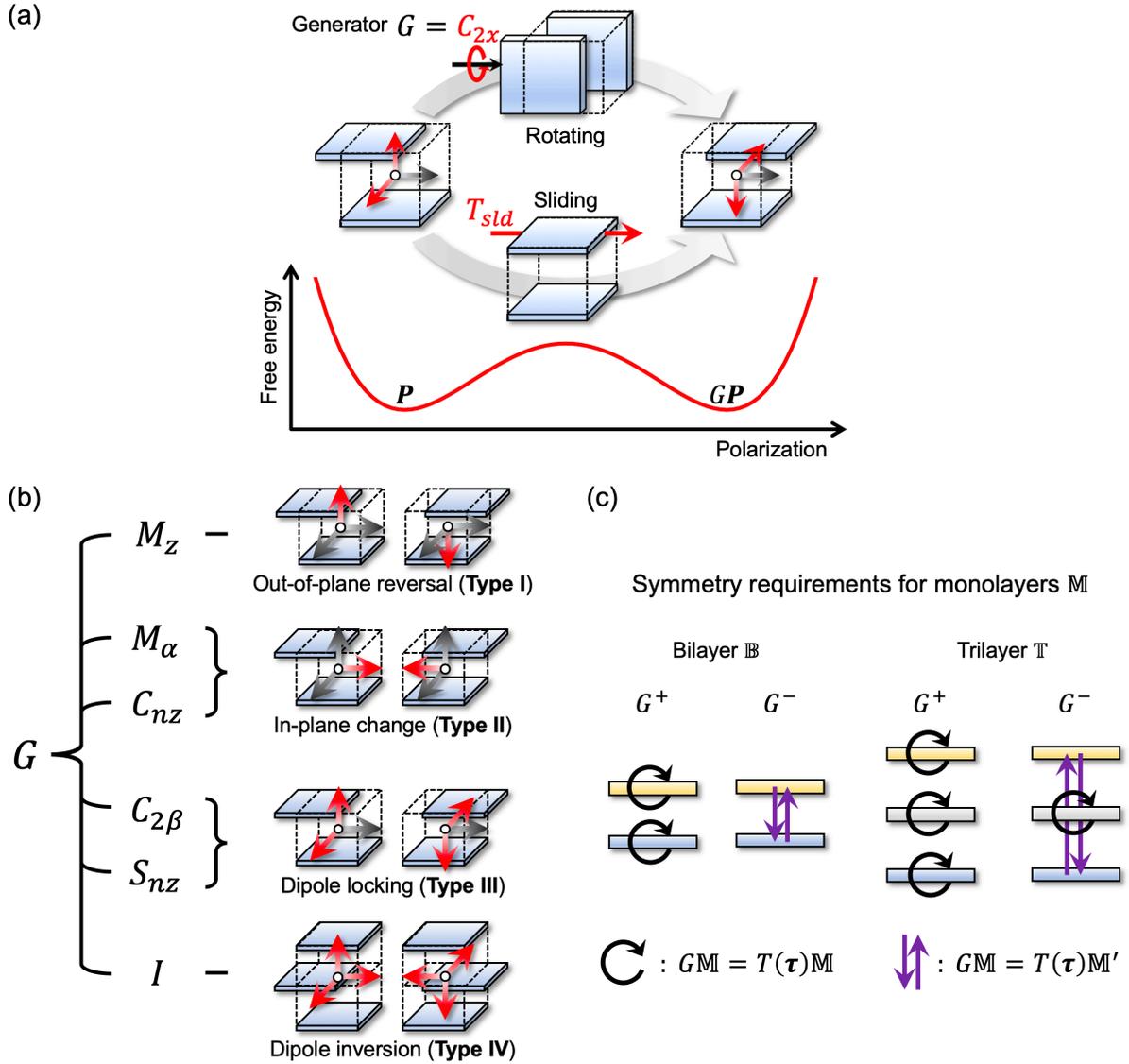

**FIG. 1.** Symmetry principle of slidetronics. (a) Schematic illustrating two equivalent approaches to relating the initial and final layer configurations in sliding FE. Sliding-induced polarization switching $\boldsymbol{P} \xrightarrow{T_{sld}} \boldsymbol{P}'$ (bottom arrow) can also be achieved by applying an appropriate point-group operator $G$ (e.g., $C_{2x}$ in this example) to the entire system (top arrow), such that $\boldsymbol{P}' = G\boldsymbol{P}$. The initial and final structures correspond to two stable phases on the free-energy surface. (b) Classification scheme of slidetronics based on the generators $G$ in Eq. (2). (c) Symmetry conditions for the constituent monolayers in a bilayer $\mathbb{B}$ (left) and in a trilayer $\mathbb{T}$ (right). Each arrow represents the transformation of one constituent layer into another (or itself) as induced by the generator $G$.

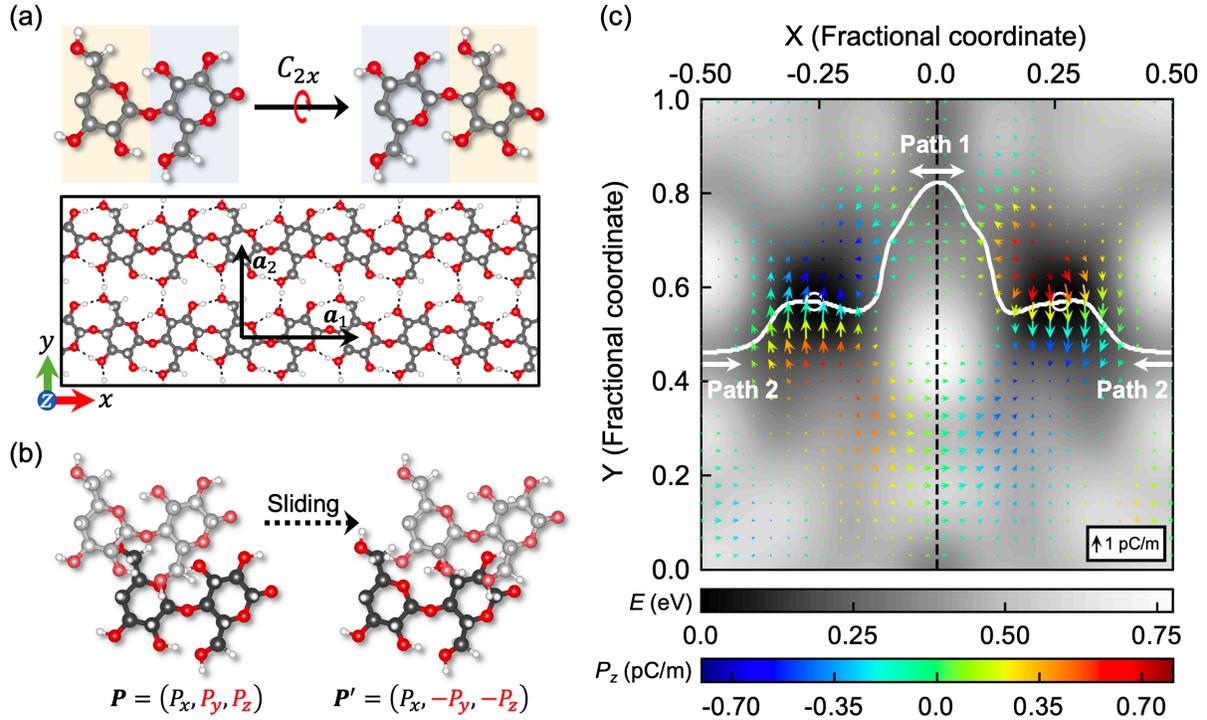

**FIG. 2.** Interlocked dipoles in cellulose bilayer. (a) Nonsymmorphic symmetry of a cellulose monolayer $\mathbb{M}$. The atomic structure remains unchanged under a 180° rotation around the $x$-axis followed by a translation of half the lattice vector $\boldsymbol{a}_1$. (b) Illustration of type-III sliding FE in a cellulose bilayer $\mathbb{B}$. (c) DFT total energy $E$ of the bilayer $\mathbb{B}$ as a function of the fractional coordinates X and Y in the translation vector $\boldsymbol{\tau}_{xy}^t = \mathrm{X}\boldsymbol{a}_1 + \mathrm{Y}\boldsymbol{a}_2$, showing two low-energy phases marked by open circles. The bilayer polarization is also plotted on the same X–Y plane. Since $P_x$ is not switchable (and thus cannot be determined), we instead plot the change in $P_x$ induced by interlayer interactions. The in-plane polarization components are indicated by arrows, while the out-of-plane component $P_z$ is represented by the color of the arrows.

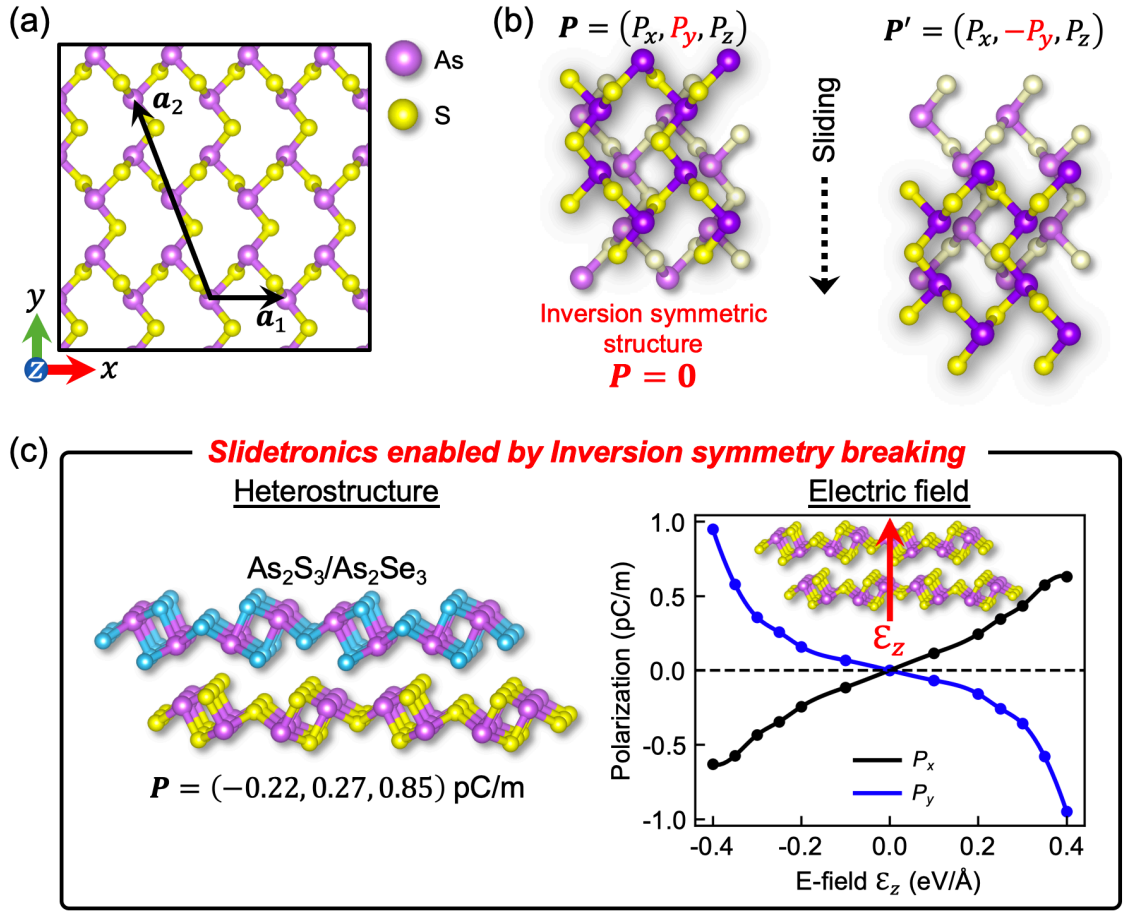

**FIG. 3.** Type-II slidetronics enabled by inversion symmetry breaking. (a) Monolayer structure of As$_2$S$_3$ ($\mathbb{M}$). (b) Inversion-symmetric bilayer configuration $\mathbb{B} = \mathbb{M} + T(\boldsymbol{\tau}_{xy}^t + \boldsymbol{\tau}_z^t)(I\mathbb{M})$ (left). (c) Two approaches to breaking the inversion symmetry of the homobilayer $\mathbb{B}$: (i) constructing a heterobilayer consisting of As$_2$S$_3$ and As$_2$Se$_3$ (left) or (ii) applying a vertical electric field $\mathcal{E}_z$ to the homobilayer $\mathbb{B}$ (right). The relationship between in-plane polarization components and $\mathcal{E}_z$ is shown. Once inversion symmetry is broken, interlayer sliding in (b) enables type-II polarization switching, $(P_x, P_y, P_z) \longrightarrow (P_x, -P_y, P_z)$.

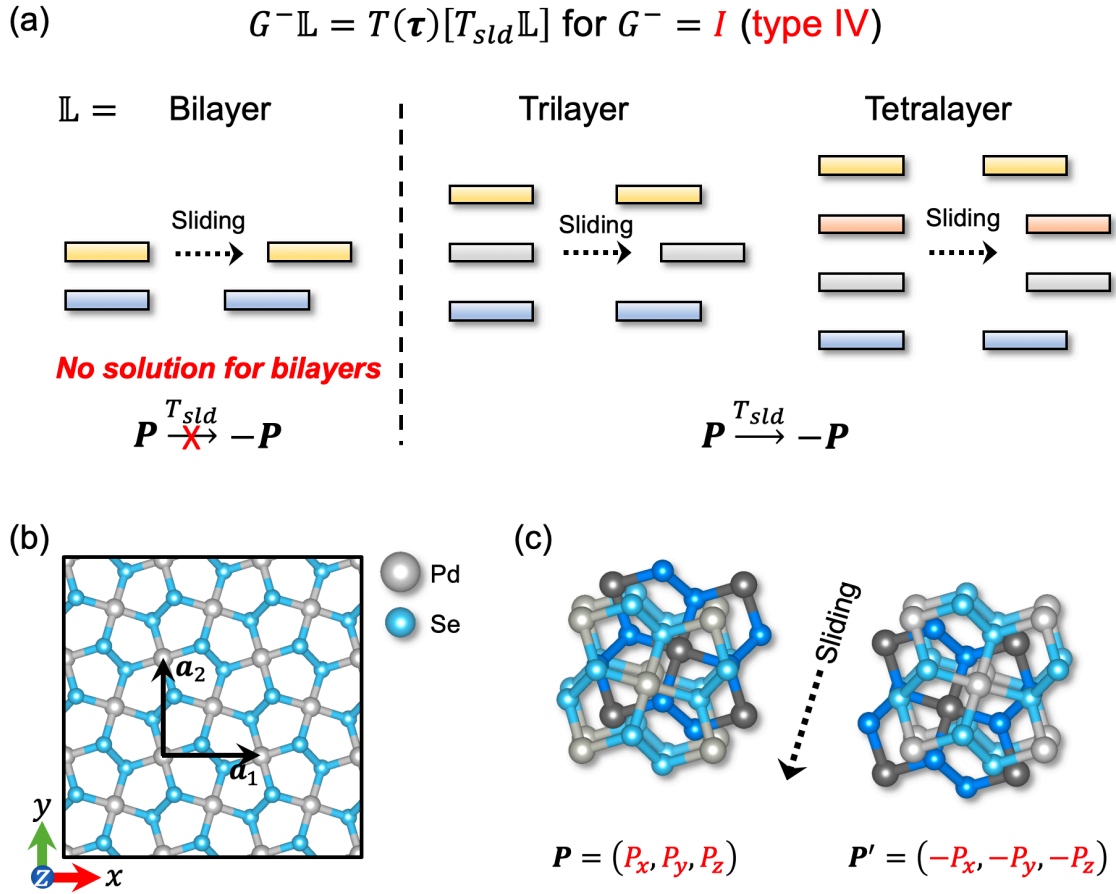

**FIG. 4.** Sliding-induced dipole inversion. (a) Schematic illustrating that a minimum of three layers is required for sliding-induced compete polarization inversion (type IV). (b–c) Demonstration of type-IV slidetronics using a PdSe$_2$ trilayer. (b) Monolayer structure of PdSe$_2$ ($\mathbb{M}$). (c) Type-IV sliding FE in a trilayer configuration, expressed as $\mathbb{T} = \mathbb{M} + T(\boldsymbol{\tau}_{xy}^m + \boldsymbol{\tau}_z^m)[C_{4z}\mathbb{M}] + T(\boldsymbol{\tau}_{xy}^t + \boldsymbol{\tau}_z^t)\mathbb{M}$.

# Supplemental Material

## Supplemental discussion

### Section S1: Slidetronics in the As₂S₃ homo-trilayer

We discuss sliding ferroelectricity (FE) in the $As_2S_3$ homo-trilayer system $\mathbb{T} = \mathbb{M} + T(\boldsymbol{\tau}_{xy}^m + \boldsymbol{\tau}_z^m)[I\mathbb{M}] + T(\boldsymbol{\tau}_z^t)\mathbb{M}$, which can be obtained from exfoliating bulk $As_2S_3$ (i.e., $\mathbb{M}_b = \mathbb{M}_t = \mathbb{M}$ and $\mathbb{M}_m = I\mathbb{M}$ for the inversion operator $I$). The most stable stacking order is shown in Supplemental Fig. S1. Starting with the stable 2D phase, we identify all other equally stable, sliding-switchable phases with distinct electric polarizations through the following steps:

**1. Generator selection:** We select all possible generators $G$ from the set $\{G\}$ in Eq. (2) that satisfy the conditions in Eqs. (7a) and (8a) (see Methods). For the $As_2S_3$ trilayer $\mathbb{T}$, there are three available generators: $G \in \{M_y, M_z, C_{2x}\}$. For $G^+ = M_y$, we obtain $\boldsymbol{\tau}^{bb} = \boldsymbol{\tau}^{mm} = \boldsymbol{\tau}^{tt} = \mathbf{0}$ from Eq. (7a). For $G^- = M_z$ and $C_{2x}$, we have $\boldsymbol{\tau}^{bt} = \boldsymbol{\tau}^{tb} = \frac{\boldsymbol{a}_2}{2}$ and $\boldsymbol{\tau}^{mm} = -\frac{\boldsymbol{a}_2}{2}$ from Eq. (8a) [Fig. 3(a)].

**2. Sliding vector calculation:** For each selected generator $G$, we calculate the sliding vectors $\boldsymbol{\tau}_{sld}^m(G)$ and $\boldsymbol{\tau}_{sld}^t(G)$ for the middle and top layers. Using Eqs. (7b) and (7c), we find the sliding vectors for $G^+ = M_y$ as $\boldsymbol{\tau}_{sld}^m(M_y) = M_y\boldsymbol{\tau}_{xy}^m - \boldsymbol{\tau}_{xy}^m = -2\boldsymbol{\tau}_y^m$ and $\boldsymbol{\tau}_{sld}^t(M_y) = \mathbf{0}$. From Eqs. (8b) and (8c), we find the same sliding vectors for $G^- = C_{2x}$, i.e., $\boldsymbol{\tau}_{sld}^m(C_{2x}) = C_{2x}\boldsymbol{\tau}_{xy}^m - \boldsymbol{\tau}_{xy}^m - \boldsymbol{a}_2 = -2\boldsymbol{\tau}_y^m$ and $\boldsymbol{\tau}_{sld}^t(C_{2x}) = \mathbf{0}$. For $G^- = M_z$, however, we obtain zero sliding vectors, $\boldsymbol{\tau}_{sld}^m(M_z) = M_z\boldsymbol{\tau}_{xy}^m - \boldsymbol{\tau}_{xy}^m - M_z\boldsymbol{\tau}_{xy}^t - \boldsymbol{a}_2 = \mathbf{0}$ and $\boldsymbol{\tau}_{sld}^t(M_z) = -M_z\boldsymbol{\tau}_{xy}^t - \boldsymbol{\tau}_{xy}^t = -2\boldsymbol{\tau}_{xy}^t = \mathbf{0}$.

**3. Final generator selection:** We retain only the generators with nonzero sliding vectors, which are $M_y$ and $C_{2x}$ for the $As_2S_3$ trilayer $\mathbb{T}$. If multiple generators have the same interlayer sliding vectors, we keep only one of them. Since $\boldsymbol{\tau}_{sld}^m(M_y) = \boldsymbol{\tau}_{sld}^m(C_{2x})$ and $\boldsymbol{\tau}_{sld}^t(M_y) = \boldsymbol{\tau}_{sld}^t(C_{2x})$ in this example, we obtain $M_y\boldsymbol{P} = C_{2x}\boldsymbol{P}$, indicating that $P_z = -P_z$ and thus $P_z = 0$. Without loss of generality, we choose $M_y$.

**4. Polarization transformation:** Each remaining generator $G$ corresponds to a trilayer phase $T_{sld}(G)\mathbb{T}$ with a polarization $\boldsymbol{P}' = G\boldsymbol{P}$. In our case with $\mathbb{T} = \mathbb{M} + T(\boldsymbol{\tau}_{xy}^m + \boldsymbol{\tau}_z^m)[I\mathbb{M}] + T(\boldsymbol{\tau}_z^t)\mathbb{M}$, we have a single generator $G = M_y$ and the associated sliding vectors $\boldsymbol{\tau}_{sld}^m = -2\boldsymbol{\tau}_y^m$ and $\boldsymbol{\tau}_{sld}^t = \boldsymbol{0}$. Therefore, the sliding operation induces a polarization transformation: $(P_x, P_y, P_z) \xrightarrow{M_y} (P_x, -P_y, P_z)$. Our DFT calculations yield $P_y = 1.76\ pC/m$, while $P_z = 0$ due to $M_z$ mirror symmetry. However, $P_x$ cannot be determined because $P_x$ is not switchable.

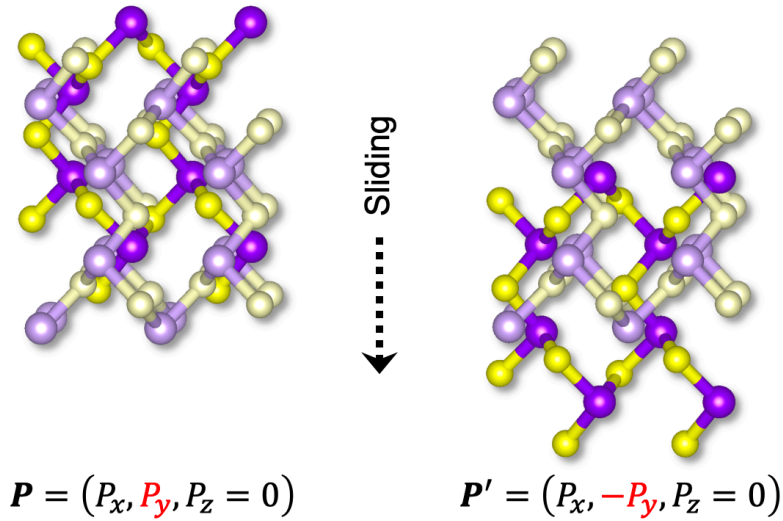

$$\boldsymbol{P} = (P_x, {\color{red}P_y}, P_z = 0) \qquad \boldsymbol{P}' = (P_x, {\color{red}-P_y}, P_z = 0)$$

**FIG. S1.** Illustration of sliding FE in the As₂S₃ homo-trilayer.

**Section S2: Absence of sliding FE in the differently stacked As₂S₃ homobilayer**

We consider a stacking configuration, $\mathbb{B}' = \mathbb{M} + T(\boldsymbol{\tau}_{xy}^t + \boldsymbol{\tau}_z^t)\mathbb{M}$, of an As₂S₃ homobilayer. In contrast to the bilayer $\mathbb{B}$ discussed in the main text, where $\mathbb{M}_b = \mathbb{M}$ and $\mathbb{M}_t = I\mathbb{M}$, here both layers share the same crystal structure, i.e., $\mathbb{M}_b = \mathbb{M}_t = \mathbb{M}$. To satisfy the conditions in Eq. (3a) or Eq. (4a), the generator $G$ should be chosen from the symmetry operations of the monolayer $\mathbb{M}$. This restricts the available generators to $G \in \{M_y, M_z, C_{2x}\}$.

Let us first consider $G^- = C_{2x}$, which satisfies Eq. (4a) with $\boldsymbol{\tau}^{bt} = \boldsymbol{\tau}^{tb} = \frac{\boldsymbol{a}_2}{2}$. From Eq. (4b), the sliding vector is given by $\boldsymbol{\tau}_{sld}^t = -C_{2x}\boldsymbol{\tau}_{xy}^t - \boldsymbol{\tau}_{xy}^t = -2\boldsymbol{\tau}_x^t$, where $\boldsymbol{\tau}_x^t$ is the translation vector along the $x$-axis. However, due to the symmetry properties of the monolayer $\mathbb{M}$, it can be shown that in the most stable configuration of $\mathbb{B}'$, the translation vector $\boldsymbol{\tau}_{xy}^t$ lies along the $y$-axis, i.e., $\boldsymbol{\tau}_{xy}^t = \boldsymbol{\tau}_y^t$ and $\boldsymbol{\tau}_x^t = \mathbf{0}$. This result is confirmed by DFT calculations (Supplemental Fig. S2). Therefore, $\boldsymbol{\tau}_{sld}^t = -2\boldsymbol{\tau}_x^t = \mathbf{0}$ for $G^- = C_{2x}$, or equivalently, $C_{2x}\boldsymbol{P} = \boldsymbol{P}$, leading to $P_y = P_z = 0$, while $P_x \neq 0$. Given the available generators $G \in \{M_y, M_z, C_{2x}\}$, it is impossible to reverse the sign of $P_x$, thereby proving that sliding FE cannot occur in the bilayer configuration $\mathbb{B}'$.

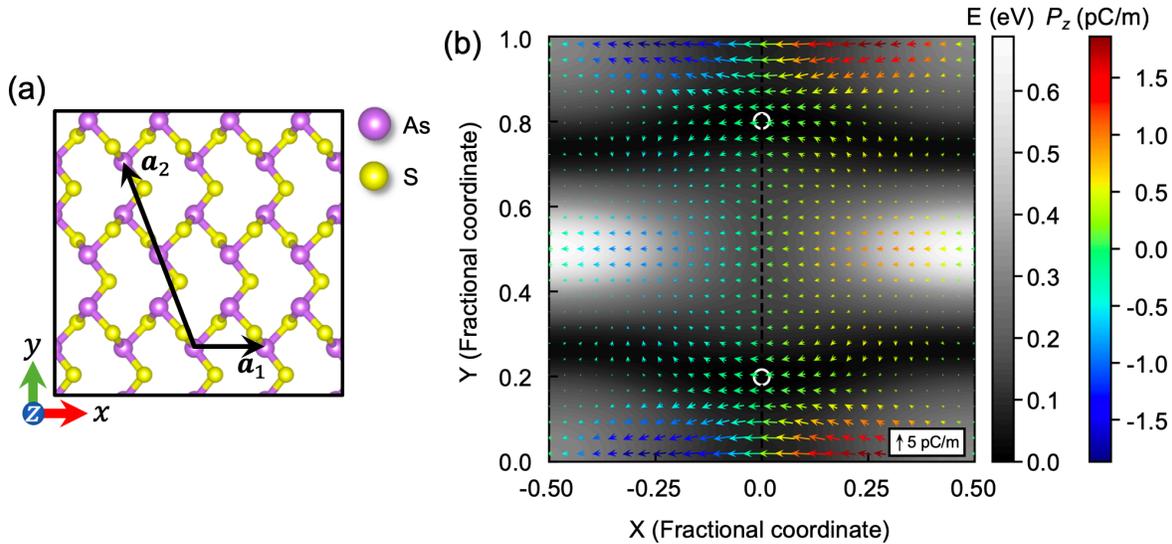

**FIG. S2.** (a) Monolayer structure of $As_2S_3$ ($\mathbb{M}$). (b) DFT total energy of the homobilayer $\mathbb{B}' = \mathbb{M} + T(\boldsymbol{\tau}_{xy}^t + \boldsymbol{\tau}_z^t)\mathbb{M}$ as a function of the fractional coordinates X and Y in the translation vector $\boldsymbol{\tau}_{xy}^t = X\boldsymbol{a}_1 + Y(\boldsymbol{a}_1 + \boldsymbol{a}_2)$, showing two low-energy phases marked by open circles. The bilayer polarization is also plotted on the same X–Y plane. Since $P_x$ is not switchable (and thus cannot be determined), we instead plot the change in $P_x$ induced by interlayer interactions. The in-plane polarization components are indicated by arrows, while the out-of-plane component is represented by the color of the arrows.

### Section S3: Type-II sliding FE in the differently stacked $As_2S_3/As_2Se_3$ heterobilayer

Sliding FE can be achieved in a heterobilayer configuration, denoted $\mathbb{B}'_{S/Se} = \mathbb{M}_S + T(\boldsymbol{\tau}_{xy}^t + \boldsymbol{\tau}_z^t)\mathbb{M}_{Se}$, composed of $As_2S_3$ ($\mathbb{M}_S$) and $As_2Se_3$ ($\mathbb{M}_{Se}$). Unlike the homobilayer $\mathbb{B}'$, where the top and bottom layers are identical and preserve $C_{2x}$ symmetry, this heterobilayer breaks $C_{2x}$ symmetry due to the different materials in each layer. This symmetry breaking induces nonzero polarization components, $P_y = 1.76 \ pC/m$ and $P_z = 0.78 \ pC/m$ (Supplemental Fig. S3).

In this heterobilayer, the generators $G^- = M_z$ and $C_{2x}$ fail to satisfy the condition in Eq. (4a). However, when using the generator $G^+ = M_y$, the top-layer sliding with $\boldsymbol{\tau}_{sld}^t = -2\boldsymbol{\tau}_y^t$ gives rise to type-II sliding FE, characterized by the sign change of $P_y$. The energy barrier associated with this sliding FE is $E_b = 2.2$ meV/Å$^2$.

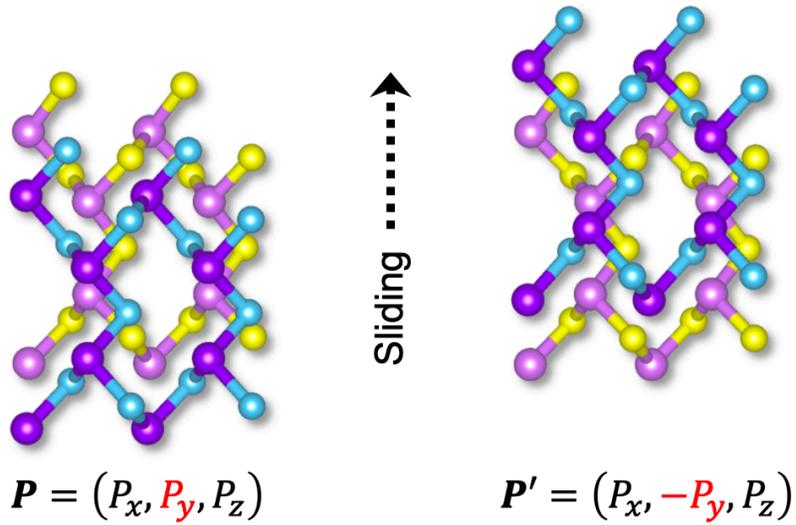

$$\boldsymbol{P} = \left(P_x, \textcolor{red}{P_y}, P_z\right) \qquad\qquad \boldsymbol{P'} = \left(P_x, \textcolor{red}{-P_y}, P_z\right)$$

**FIG. S3.** Illustration of type-II sliding FE in the As$_2$S$_3$/As$_2$Se$_3$ heterobilayer $\mathbb{B}'_{S/Se}$.